\documentstyle[prl,preprint,aps,epsf]{revtex}

\begin{document}
\bibliographystyle{plain}
\title{General Criterion for the existence of Supertube and BIon 
in Curved Target Space}
\author{D. K. Park$^{1,3}$\footnote{e-mail:
dkpark@hep.kyungnam.ac.kr},
S. Tamaryan$^{2,4}$\footnote{e-mail: sayat@moon.yerphi.am} and
H. J. W. M\"{u}ller-Kirsten$^{3}$\footnote{e-mail:
mueller1@physik.uni-kl.de}
}
\address{
1.Department of Physics, Kyungnam University, Masan, 631-701, Korea\\
2.Theory Department, Yerevan Physics Institute, Yerevan-36, 375036, Armenia\\
3.Department of  Physics, University of Kaiserslautern, 
  67653 Kaiserslautern, Germany\\
4.Department of Mathematical Physics, NUIM, Maynooth, Ireland 
}
\date{\today}
\maketitle

\begin{abstract}
The supertube and BIon spike solutions are examined in a  general 
curved target space. The criteria for the existence of these solutions
are explicitly derived. Also the equation which the general BIon solution
should satisfy is derived.

\end{abstract}
\maketitle
\newpage
Although there are many kinds of bound states, considerable attention
has  recently been
paid to the  supertube\cite{mat01}, which is a bound state of D2-brane, 
F-string, and D0-brane. The supertube is a BPS object and preserves
$(1/4)$ of supersymmetry. It is supported against collapse by the 
Poynting 
angular momentum which is generated by the worldvolume electromagnetic 
fields.

The physical implications
 for the upper bound of  angular momentum have been  investigated
on the supergravity side\cite{empa01} and from
an  $AdS$/CFT setting\cite{gre00}.
From the viewpoint of supergravity the maximum angular momentum is related
to a  global violation of causality. From the $AdS/CFT$
viewpoint  it is related
to the maximum size of the particle which is a `blown-up' version
 of the massless
particle {\it via} Myers'  effect\cite{myers99}. Other aspects of the maximal
angular momentum have been summarized in Ref.\cite{town02}.

The supertube has recently
found  further  generalizations,  and its  T-dual, 
TST-dual, and M-theoretical variants have been 
constructed\cite{bak01,cho01,bak02,mat02,cho02,mat02-2,hyaku02,tama02,park03}.
The interesting feature  is that  supersymmetry
 is not broken, although the 
shape of the cross section is arbitrary\cite{mat02}. This property 
has been  
re-examined in the context of M-theory\cite{hyaku02}, which yields a 
unified description for the supertube, D-Helix\cite{cho01}, and 
supercurve\cite{mat02-2} in terms of M-Ribbons. 
Another interesting aspect  of 
the supertube is the fact that it can be deformed into a  
D2-anti-D2 system, which 
has been  explored in the context of  Matrix theory\cite{bak01,bak02}
and abelian Born-Infeld theory\cite{mat02,cho02}.
 From this aspect the 
supertube can be understood as a miraculous separation of D2-anti-D2 branes.
The D2-anti-D2 configuration has also been  extended to the 
spherical sphere\footnote{In $S^2$ topology we cannot 
usually define  the 
electromagnetic fields smoothly due to the poles. Thus, one has  to ignore
the problem arising from the  topology  
in this case\cite{pri}.}\cite{park03} and
D3-branes\cite{tama02}.

Here  we   consider general criteria  for the existence of 
the supertube and BIon solutions in a  curved target space.   
This was  initially
motivated from the observation that we cannot find any explicit reason
in the original paper\cite{mat01}
why a flat target space should be  essential in  forming the supertube. 
In the following 
 we  derive this criterion explicitly,  and also 
an equation which the general BIon solution would have to  satisfy.
Thus our aim is 
 to obtain supertube-like and BIon-like solutions
in  curved spacetime. To this end   
we use  a simple observation  valid in flat spacetime.
As a starting point let us recall  the  energy density for the  supertube
given in Ref.\cite{mat01} (setting $2\pi\alpha^{\prime}=1$
and $2\pi g_s=1$)
\begin{equation}
\label{ST0}
{\cal H}_{ST} = \frac{1}{R}
\sqrt{(\Pi^2 + R^2) (B^2 + R^2 + R^2 R_x^2)},
\end{equation}
where $R_x \equiv \partial_x R$ and we chose  $\partial_{\varphi} R = 0$
for simplicity. Of course,  $\Pi$ is the conjugate momentum
of the electric field,  and $B$  is 
the worldvolume magnetic field. If $R_x = 0$, the energy density 
simply reduces to 
\begin{equation}
\label{ST1}
{\cal H}_{ST} = \frac{1}{R}
\sqrt{(\Pi^2 + R^2) (B^2 + R^2)}.
\end{equation}
It is instructive  to re-express ${\cal H}_{ST}$ in the 
following form with inequality:
\begin{equation}
\label{ST2}
{\cal H}_{ST} = 
\sqrt{(\Pi + B)^2 + \left( \frac{B}{R} \Pi - R \right)^2}
\geq \Pi + B.
\end{equation} 
With  Eq.(\ref{ST2}) one can easily understand that 
the inequality  (\ref{ST2}) is saturated by the  condition 
$R^2 = \Pi B$,  and the corresponding BPS energy of the supertube is
${\cal H}_{ST}^{BPS} = \Pi + B$. Since the supertube does not carry 
D2-brane charge, it can be regarded as a `blown-up' form  of the type IIA
superstring which carries D0-brane charge. Furthermore, Ref.\cite{mat01}
has shown that this   configuration  preserves  $(1/4)$-supersymmetry.

Ref.\cite{mat01} has also obtained the BIon spike solution which also
preserves $(1/4)$-supersymmetry. We first demonstrate
 how the BIon solution
can be reproduced from using a  simple inequality.
When $R_x \neq 0$, the energy density (\ref{ST0}) cannot be
expressed in a  form,  from which  an inequality can  easily
be constructed.
This difficulty is overcome  if we impose the
condition   $B = b R R_x$. In this case
the energy density (\ref{ST0}) can be  expressed as follows:
\begin{equation}
\label{ST3}
{\cal H}_{ST} =
\sqrt{ \left( \Pi + \frac{\sqrt{1 + b^2}}{b} B \right)^2 +
       \left( \frac{\frac{\sqrt{1 + b^2}}{b} B}{R} \Pi - R \right)^2
     } \geq \Pi + \frac{\sqrt{1 + b^2}}{b} B.
\end{equation} 
One should note that this is not a BPS-type inequality because the
r.h.s of Eq.(\ref{ST3}) is dependent on $R$ {\it via} the magnetic
field. In spite of this, however, we can show that this inequality gives
a BIon spike solution at the saturation point.

The inequality is saturated for
$R^2 = \sqrt{1 + b^2} B \Pi / b$, and the corresponding energy density is
\begin{equation}
\label{ST4}
{\cal H}_{ST}^{MBPS} = \Pi + \frac{\sqrt{1 + b^2}}{b} B,
\end{equation}
where $MBPS$ stands for modified BPS. 
The first term in ${\cal H}_{ST}^{MBPS}$ is interpreted as a 
contribution from the F-string and the second term
as a contribution  from the D2-anti-D2
system\cite{cho02}.
Although we obtained 
${\cal H}_{ST}^{MBPS}$ as a saturation condition of the inequality, 
we cannot guarantee that the configuration we obtained is a classical 
solution in view  of the $R$-dependence of the r.h.s. of Eq.(\ref{ST3}). 
Thus, we should check this  explicitly by
using the equations of motion. It is a  simple matter to show that
the saturation condition $R^2 = \sqrt{1 + b^2} B \Pi / b$ with 
$B = b R R_x$ yields 
\begin{equation}
\label{ST5} 
R = C e^{\frac{x}{E_0}},
\hspace{1.0cm}
E_0 = \Pi \sqrt{1 + b^2},
\end{equation}
which is nothing but the BIon spike solution.
This means, the inequality (\ref{ST3}) also solves the equations of motion.

We  use this simple observation to construct   supertube-like and
BIon-like solutions in the 
curved target spacetime.
For this we  consider the usual
D2-worldvolume Lagrangian ${\cal L} = - \sqrt{-det(g + F)}$ in  target 
space (conventions as stated above)
\begin{equation}
\label{target1}
ds^2 = -W_1(R) dT^2 + W_3(R) dX^2
 + W_2(R) dR^2 + W_4(R) R^2 d\Phi^2 + ds^2(E^6),
\end{equation}
where $W_1(R)$, $W_2(R)$, $W_3(R)$, and $W_4(R)$ are  arbitrary functions 
of $R$. We first  consider the case of  $W_3 = W_4 = 1$ 
for simplicity. 
We  describe the result for the case
with  arbitrary  $W_3$ and $W_4$ at the 
end of the  paper.  

Defining 
the worldvolume coordinates  $(t, x, \varphi)$ as 
\begin{equation}
\label{wvolume1}
t=T\, ,\qquad x = X \, ,\qquad \varphi=\Phi\, 
\end{equation}
and choosing the BI 2-form field strength as 
\begin{equation}
\label{BIform1}
F = E dt \wedge dx + B dx \wedge d\varphi,
\end{equation}
it is straightforward to compute the Lagrangian ${\cal L}$
\begin{equation}
\label{lag1}
{\cal L} \equiv -\bigtriangleup_1 = 
- \sqrt{(R^2 + W_2 R_{\varphi}^2) (W_1 - E^2) + W_1 B^2 + W_1 W_2 R_x^2 R^2}
\end{equation}
where $R_{\xi} = \partial_{\xi} R$. 
We   note  the factor $W_1 - E^2$ in the square root
of Eq.(\ref{lag1}). If we start with a  flat target space, {\it i.e.}
$W_1 = W_2 = 1$, the critical electric field would be $E = 1$, which 
indicates that the  velocity of the helix of Ref.\cite{mat01}
 is the  velocity of light. Thus, the factor
$W_1 - E^2$ indicates that the  velocity 
of light is changed to $\sqrt{W_1}$
in this background. This can  also be  realized directly from the target
spacetime metric (\ref{target1}).

The conjugate  momentum of the electric field  $E$, is
\begin{equation}
\label{displace1}
\Pi = \frac{\partial {\cal L}}{\partial E} =
\frac{R^2 E}{\bigtriangleup_1},
\end{equation}
and Eq.(\ref{displace1}) enables us to re-express the electric field $E$
in terms of $\Pi$ and $B$:
\begin{equation}
\label{elect1}
E = \frac{\Pi}{R}
\sqrt{\frac{W_1 [R^2 + B^2 + W_2 R_x^2 R^2]}{\Pi^2 + R^2}}.
\end{equation}
Eq.(\ref{elect1}) allows us to rewrite the
 energy density in the well-known form
\begin{equation}
\label{edens1}
{\cal H}_{W} = \Pi E + \bigtriangleup_1
= \frac{1}{R}
\sqrt{W_1 (\Pi^2 + R^2) (R^2 + B^2 + W_2 R_x^2 R^2)}.
\end{equation}

We now  consider first the case
 $R_x = 0$, which corresponds to the case of the 
supertube in  the flat spacetime background. In this case the energy
density becomes
\begin{eqnarray}
\label{edens2}
{\cal H}_{W}&=& \frac{1}{R}
\sqrt{W_1 (\Pi^2 + R^2) (B^2 + R^2)}    \\  \nonumber
&=&\sqrt{W_1 (\Pi + B)^2 + W_1 
      \left( \frac{B}{R} \Pi - R \right)^2}
      \geq \sqrt{W_1} (\Pi + B).
\end{eqnarray}
It is interesting to note that the energy density ${\cal H}_{W}$ is 
always expressible as an inequality regardless of $W_1$ 
and $W_2$.
Thus, the inequality (\ref{edens2}) is saturated at 
\begin{equation}
\label{satu1}
R^2 = \Pi B
\end{equation}
and the corresponding energy density becomes
\begin{equation}
\label{bpsene1}
{\cal H}_{W}^{MBPS} = \sqrt{W_1} (\Pi + B).
\end{equation}

The saturation condition (\ref{satu1}) forces
 $E$ in Eq.(\ref{elect1}) to 
assume  its critical value $E = \sqrt{W_1}$, as expected. The factor
$\sqrt{W_1}$ in Eq.(\ref{bpsene1}) is   part of  $\sqrt{-g}$, which
is necessary for the definition of volume (or surface) element in the
curved spacetime.

Although we have obtained the configurations $R^2 = \Pi B$ and
$E = \sqrt{W_1}$, we should check whether these solve the classical
equations of motion or not, because the saturation of  the inequality
(\ref{edens2}) alone  does not guarantee this
 in view of the  $R$-dependence of the 
r.h.s. of (\ref{edens2}) {\it via} $W_1$. The equations  of motion
or constraints  with
respect to the gauge potentials are
\begin{equation}
\label{eqmotion1}
\frac{\partial \Pi}{\partial x} =
\frac{\partial H}{\partial x} =
\frac{\partial H}{\partial \varphi} = 0,
\end{equation}
where
\begin{equation}
\label{defmon1}
\Pi = \frac{R^2 E}{\bigtriangleup_1},
\hspace{1.0cm}
H = \frac{W_1 B}{\bigtriangleup_1}.
\end{equation}
One can easily show  that our configurations solve these
equations  (\ref{eqmotion1}) if and only if $W_1 = const$.
Thus the existence of the  supertube-like solution
 requires  the time-time 
component of the metric to
be equal to  that of flat space. Later we will show that these
configurations preserve $(1/4)$-supersymmetry like the
flat spacetime 
supertube. We point out that
verifying  that a constraint like $\partial H/\partial x =0$
in Eq.(\ref{eqmotion1}) is satisfied can be done
by evaluating the variation of ${\cal H}_W$ of Eq. (\ref{edens1})  
for $B\propto RR_x$.  The latter propotionality requires
precisely this constraint. Ref.\cite{mat01} mentions only
the first constraint, the Gauss law.

Next we turn to the case  $R_x \neq 0$.
 As usual, in this case, it seems
to be impossible to express ${\cal H}_{W}$ in Eq.(\ref{edens1}) as a 
form from which the the inequality can be  easily constructed. 
As shown, however, in the 
previous section, this difficulty  can be overcome
 by imposing the 
condition satisfying the constraints 
(\ref{eqmotion1}), i.e.
\begin{equation}
\label{impose1}
B = B_0 R R_x.
\end{equation}
Then the energy density in Eq.(\ref{edens1}) assumes the  
desired form:
\begin{equation}
\label{edens3}
{\cal H}_{W} = 
\sqrt{W_1 \left(\Pi + \sqrt{1 + \frac{W_2}{B_0^2}} B \right)^2
      + W_1 \left(\frac{\sqrt{1 + 
\frac{W_2}{B_0^2}} B}{R} \Pi - R
                                                              \right)^2     }
\geq \sqrt{W_1} \left( \Pi + \sqrt{1 + \frac{W_2}{B_0^2}} B \right).
\end{equation}
Again this inequality is saturated at
\begin{equation}
\label{satu2}
R^2 = \sqrt{1 + \frac{W_2}{B_0^2}} \Pi B,
\end{equation}
and the corresponding energy density is 
\begin{equation}
\label{bpsene2}
{\cal H}_{W}^{MBPS} = \sqrt{W_1}
\left[ \Pi + \sqrt{1 + \frac{W_2}{B_0^2}} B \right].
\end{equation}
We note that the saturation condition (\ref{satu2})
 with (\ref{impose1})
also requires  $E$ in Eq.(\ref{elect1}) to be $E = \sqrt{W_1}$.

Although we have obtained ${\cal H}_{W}^{MBPS}$, it is not simple to 
interpret the second term of Eq.(\ref{bpsene2}) for arbitrary $W_2$. 
However, we should note that the expression (\ref{edens3}) for 
${\cal H}_{W}$ is always possible although $B_0$ is not constant but
function of $R$. Thus, to make a correct interpretation we assume that 
$B_0$ is a function of $R$ as $B_0 = B_1 \sqrt{W_2}$. Then the 
energy density ${\cal H}_{W}^{MBPS}$ assumes the  familar form
\begin{equation}
\label{bpsene3}
{\cal H}_{W}^{MBPS} = \sqrt{W_1}
\left[ \Pi + \frac{\sqrt{1 + B_1^2}}{B_1} B \right],
\end{equation}
and the corresponding magnetic field becomes
\begin{equation}
\label{magne1}
B = B_1 \sqrt{W_2} R R_x.
\end{equation}

Although all difficulties seem to be removed, there is an additional 
problem in the general curved background. If $B_1$ in Eq.(\ref{magne1})
is a constant, the configurations we obtained here do not solve the 
equations of motion due to the factor $W_1$ in front of $B^2$ in
BI Lagrangian (\ref{lag1}). This means we need an additional technique 
to resolve this difficulty. The only method solving this problem is to 
re-define $B_1$ again as
\begin{equation}
\label{magne2}
B_1 = \frac{b}{\sqrt{(1 + b^2) W_1 - b^2}}
\end{equation}
where $b$ is some constant. Then, the equation of motion for $H$ is 
automatically
solved and the saturated energy density becomes
\begin{equation}
\label{bpsene3}
{\cal H}_{W}^{MBPS} = \sqrt{W_1} \Pi +
\frac{\sqrt{1 + b^2}}{b} W_1 B.
\end{equation}
Of course, the first term in (\ref{bpsene3}) can be interpreted as a
contribution from F-strings  and the second term
as a contribution  from the  D2-anti-D2 system.
Note that the power of $W_1$ is different in these contributions. This
means,  the energy of the D2-anti-D2 system depends  nontrivially
 on the curved spacetime background.

The remaining equations of motion are solved when the equations
\begin{equation}
\label{complicate}
R_x = \frac{\sqrt{(1 + b^2) W_1 - b^2}}{\Pi \sqrt{(1 + b^2) W_1 W_2}} R
\end{equation}
and 
\begin{equation}
\label{complicate1}
\frac{W_1^{\prime}}{2 \sqrt{W_1}} 
\left[ 1 + \frac{R^2}{\Pi^2} \left( 1 - \frac{1}{R_x} \right) \right]
+ \frac{\sqrt{W_1}}{2} W_2^{\prime} R_x (R_x - 1) = 0
\end{equation}
are satisfied, where the prime denotes the differentiation with respect to 
the argument. 
Eq.(\ref{complicate}) is just the  Gauss law and yields the
 usual BIon spike
solution in  the flat spacetime  limit.
 Eq.(\ref{complicate1}) is obtained by 
varying the action (\ref{lag1}) with respect to $R$. In  deriving 
Eq.(\ref{complicate1}) we have used Eq.(\ref{complicate}) for
reasons of  
 simple appearance. Of course, this
 radial equation is automatically solved
in  the flat spacetime  limit.

Although the BIon-like solution $R_x$ 
can be obtained from Eq.(\ref{complicate}) for some  particular
$W_1$ and $W_2$, it is a  different matter whether the solution preserves
the expected partial supersymmetry or not. 
This means the condition for supersymmetry preservation may give another
constraint.
To examine this,  we consider
the Killing equation $\Gamma \epsilon = \epsilon$ where $\Gamma$ is a 
projection matrix appearing in the `$\kappa$-symmetry' transformation.
Then we obtain two conditions for the preservation of supersymmetry
\begin{eqnarray}
\label{susy1}
& &\left[ \sqrt{W_1} \Gamma_{TX} \Gamma_{\natural} + E \right] 
                                       \epsilon_0 = 0, \\ \nonumber
& & \left[\sqrt{W_2} R R_x \Gamma_{T R \Phi} + B \Gamma_{T}
\Gamma_{\natural} \right] \epsilon_0 = \frac{\bigtriangleup_1}{\sqrt{W_1}}
                                       \epsilon_0,
\end{eqnarray} 
where the  conventions  of Ref.\cite{mat01} have been  used.
The first equation of (\ref{susy1}) is consistent with  our 
constraint $E = \pm \sqrt{W_1}$. However, the second equation of 
(\ref{susy1}) becomes in  our case
\begin{equation}
\label{susy2}
\left[\Gamma_{T R \Phi} + \frac{b}{\sqrt{(1 + b^2) W_1 - b^2}}
\Gamma_{T} \Gamma_{\natural} \right] \epsilon_0 = 
\frac{\sqrt{(1 + b^2) W_1}}{\sqrt{(1 + b^2) W_1 - b^2}} \epsilon_0.
\end{equation}
To assign a proper physical meaning the only possible way is to 
choose $W_1 = 1$, which reduces  Eq.(\ref{susy2}) to  the 
well-known form $(\Gamma_{T R \Phi} + b \Gamma_{T} \Gamma_{\natural})
\epsilon_0 = \sqrt{(1 + b^2)} \epsilon_0$. Hence, again we obtained 
the condition $W_1 = 1$ for the existence of the BIon-like solution.

Thus the  supersymmetry condition
 reduces  equation (\ref{complicate}) to the
simple form
\begin{equation}
\label{final} 
R_x = \frac{1}{\Pi \sqrt{(1 + b^2) W_2}} R.
\end{equation}  
Furthermore, the radial equation (\ref{complicate1}) becomes simply
$W_2^{\prime} R_x (R_x - 1) = 0$. As we commented in Ref. \cite{park03}
the radial equation is not needed to be an equation of motion if one
 starts with a  static theory from the
very  beginning because in this case $R$ is not a
dynamical variable. Thus we encounter a 
 somewhat subtle situation. If we do not 
regard  the radial equation  as an equation of motion, we have a 
BIon-like solution obtained by solving Eq.(\ref{final}), which 
preserves $(1/4)$-supersymmetry. But  if we do
regard the radial equation
as an equation of motion, this  fixes either $W_2 = const$ or $R_x = 1$.
Since in this case the first choice makes  spacetime  flat,  once
again, we think,  the second choice is the more interesting. 

However, this situation 
is drastically changed when we have arbitrary $W_i (i = 1, 2, 3, 4)$.
In this case the condition for the 
preservation of the $(1/4)$-supersymmetry also generates the constraint
$W_1 = 1$.  In the case of  the supertube,
 however, the equation of motion yields
the further  constraint $W_3 = 1$ on the  target spacetime.
 In fact, this is 
an expected result because we know  that the cross section of 
the supertube can be arbitrary. This means,
 we have   freedom in  choosing 
$W_2$ and $W_4$. The physical meaning of $W_3 = 1$ is that the axis of 
the supertube should be straight. This fact has  also  been confirmed 
in Ref.\cite{hyaku02} from the viewpoint of M-theory. 

For the BIon case the equations corresponding to Eqs.(\ref{complicate}) 
and (\ref{complicate1}) are 
\begin{eqnarray}
\label{full1}
& &
\hspace{3.0cm}
R_x = \frac{1}{\Pi} 
      \sqrt{\frac{W_3 W_4 [(1 + b^2) W_1 - b^2]}{(1 + b^2) W_1 W_2}} R
                                                         \\  \nonumber 
& & 
\frac{W_1^{\prime}}{2 \sqrt{W_1 W_3}}
\left[ 1 + \frac{W_3 W_4}{\Pi^2} R^2 \left( 1 - \frac{1}{R_x} \right) \right]
+ \frac{1}{2} \sqrt{\frac{W_1}{W_3}}
\left[ W_3^{\prime} + R_x (R_x - 1)
      \left( W_2^{\prime} + \frac{W_2}{W_4} W_4^{\prime}\right) \right] = 0.
\end{eqnarray}
Thus the condition $W_1 = 1$ makes these equations to be more treatable
form
\begin{eqnarray}
\label{full2}
R_x&=& \frac{1}{\Pi} 
\sqrt{\frac{W_3 W_4}{(1 + b^2) W_2}} R,
                                         \\   \nonumber
W_3^{\prime}&+&R_x (R_x - 1) 
      \left( W_2^{\prime} + \frac{W_2}{W_4} W_4^{\prime}\right) = 0.
\end{eqnarray}
Although, therefore, we regard the radial constraint as an equation of motion,
it cannot fix all of functions $W_i$ in this case. Thus, the solution of 
Eq.(\ref{full2}) for the fixed $W_2$, $W_3$, and $W_4$ is a general BIon
spike solution in the curved target space.

{\bf Acknowledgement:} DKP acknowledges support by 
the  Korea Research Foundation under 
Grant (KRF-2002-015-CP0063) and the Deutsche Forschungsgemeinschaft
(Germany). ST acknowledges support from INTAS project
INTAS-00-00561.










\begin{thebibliography}{99}

%

\bibitem{mat01} D. Mateos and P. K. Townsend, {\it Supertubes}, Phys. Rev. Lett.
{\bf 87} (2001) 011602 [hep-th/0103030].

\bibitem{empa01} R. Emparan, D. Mateos  and P. K. Townsend, {\it Supergravity
Supertubes}, JHEP {\bf 0107} (2001) 011 [hep-th/0106012].

\bibitem{gre00} J. McGreevy, L. Susskind  and N. Toumbas, {\it Invasion of
Giant Gravitons from Anti de Sitter Space}, JHEP {\bf 0006} (2000) 008
[hep-th/0003075].

\bibitem{myers99} R. C. Myers, {\it Dielectric Branes}, JHEP {\bf 9912}
(1999) 022 [hep-th/9910053].

\bibitem{town02} P. K. Townsend, {\it Surprises with Angular Momentum},
[hep-th/0211008].

\bibitem{bak01} D. Bak and K. Lee, {\it Noncommutative Supersymmetric
Tubes}, Phys. Lett. {\bf B 509} (2001) 168 [hep-th/0103148].

\bibitem{cho01} J. H. Cho and P. Oh, {\it Super D-Helix}, Phys. Rev. 
{\bf D64} (2001) 106010 [hep-th/0105095].

\bibitem{bak02} D. Bak and A. Karch, {\it Supersymmetric Brane-AntiBrane
Configurations}, Nucl. Phys. {\bf B626} (2002) 165 [hep-th/0110039].

\bibitem{mat02} D. Mateos, S. Ng  and P. K. Townsend, {\it Tachyons,
Supertubes and Brane/Anti-Brane Systems}, JHEP {\bf 0203} (2002)
016 [hep-th/0112054].

\bibitem{cho02} J. H. Cho and P. Oh, {\it Elliptic supertube and a
Bogomol'nyi-Prasad-Sommerfield D2-brane-anti-D2-brane pair}, Phys. Rev. 
{\bf D65}
(2002) 121901 [hep-th/0112106].

\bibitem{mat02-2} D. Mateos, S. Ng  and P. K. Townsend, {\it Supercurves},
Phys. Lett. {\bf B538} (2002) 366 [hep-th/0204062].

\bibitem{hyaku02} Y. Hyakutake and N. Ohta, {\it Supertubes and 
Supercurves from M-Ribbons}, Phys. Lett. {\bf B539} (2002) 153
[hep-th/0204161].

\bibitem{tama02} S. Tamaryan, D. K. Park, and H. J. W. M\"{u}ller--Kirsten,
{\it Tubular D3-branes and their Dualities} [hep-th/0209239].

\bibitem{park03} D. K. Park, S. Tamaryan, and H. J. W. M\"{u}ller--Kirsten,
{\it Supersphere}, Phys. Lett. {\bf B551} (2003) 187 [hep-th/0210306].

\bibitem{pri} Private commutation from Y. Hyakutake.

\end{thebibliography}
\end{document}